\documentclass[conference]{IEEEtran}
\usepackage{cite}
\usepackage{amsmath,amssymb,amsfonts}
\usepackage{algorithmic}
\usepackage{graphicx}
\usepackage{textcomp}
\usepackage{xcolor}
\usepackage{soul}
\usepackage{enumerate}
\usepackage{comment}

\usepackage{subcaption}
\usepackage{multirow}
\usepackage{footnote}
\usepackage{tabu}
\usepackage{makecell}
\usepackage{float}
\usepackage{adjustbox}
\usepackage{booktabs}
\usepackage[inline]{enumitem}
\usepackage{lipsum}

\definecolor{boristext}{rgb}{0.1, 0.44, 0.84}
\definecolor{boriscomments}{rgb}{0.8, 0.2, 0.04}

\def\BibTeX{{\rm B\kern-.05em{\sc i\kern-.025em b}\kern-.08em
    T\kern-.1667em\lower.7ex\hbox{E}\kern-.125emX}}
    
\DeclareMathOperator*{\E}{\mathbb{E}}    
\definecolor{boristext}{rgb}{0.22, 0.44, 0.88}
\definecolor{boriscomments}{rgb}{0.88, 0.04, 0.04}
\definecolor{boristochange}{rgb}{0.2, 0.8, 0.8}

\begin{document}

\title{TXOP sharing with Coordinated Spatial Reuse in Multi-AP Cooperative IEEE 802.11be WLANs}

\author{D. Nunez$^1$, F. Wilhelmi$^2$, S. Avallone$^3$, M. Smith$^4$, B. Bellalta$^1$\\ $^1$ Wireless Networking group, Universitat Pompeu Fabra. e-mail:\{david.nunez,boris.bellalta\}@upf.edu \\
$^2$ Centre Tecnològic de Telecomunicacions de Catalunya (CTTC). e-mail: fwilhelmi@cttc.cat\\
$^3$ Department of Computer Engineering, University of Napoli Federico II. e-mail: stavallo@unina.it  \\
$^4$ CISCO systems. e-mail: mmsmith@cisco.com}

\maketitle

\begin{abstract}

 
IEEE 802.11be networks (aka Wi-Fi 7) will have to cope with new bandwidth-hungry and low-latency services such as eXtended Reality and multi-party cloud gaming. With this goal in mind, transmit opportunity (TXOP) sharing between coordinated access points (APs) may contribute to alleviating inter-AP contention, hence increasing the overall network throughput. This paper evaluates two coordinated TXOP sharing strategies: coordinated time division multiple access (c-TDMA) and coordinated-TDMA with spatial reuse (c-TDMA/SR). We show that, while c-TDMA alone does not result in any significant improvement in terms of the WLAN throughput,
it lays the groundwork to implement coordinated SR (c-SR) techniques. To evaluate the performance of c-TDMA/SR, we propose a fair scheduler able to select the best subset of parallel transmissions in WLAN deployments, as well as the appropriate power levels to be used by APs and stations (STAs), leading to maximum performance. The results obtained for c-TDMA/SR show significant throughput gains compared with c-TDMA, with values higher than 140\% in 90\% of the considered scenarios.

 
\end{abstract}

\begin{IEEEkeywords}
coordinated spatial reuse, coordinated TDMA, IEEE 802.11be, multi access point coordination, WLAN. 
\end{IEEEkeywords}



\section{Introduction}

Wi-Fi technology is rapidly evolving to meet the ever-tighter user demands. At this moment, when Wi-Fi 6 based on the IEEE 802.11ax amendment \cite{bellalta2016ieee} is already commercially available, the research efforts are focused on the development of the IEEE~802.11be (11be) amendment~\cite{draft11be}, which will be the main building block of the future Wi-Fi 7. This amendment will bring up higher data rates due to the use of higher modulation orders, large channel widths and more spatial streams, as well as a set of new features such as multi-link operation (MLO) and multi access point coordination (MAPC) \cite{AdrianGarciaSurvey,EvgenySurvey}. While MLO is currently receiving a lot of attention from the research community \cite{naribole2020simultaneous,naik2021can,lopez2021ieee,park2021latency,carrascosa2021experimental}, the number of works considering MAPC is still very low. It could be because MLO is planned for the first 11be amendment release, and MAPC will be included in the second one.


MAPC represents a paradigm shift for future Wi-Fi 7 networks and beyond. It allows the AP that wins the contention to share its transmit opportunity (TXOP) with other APs. TXOP sharing also allows the TXOP-winner AP to coordinate the available temporal, frequency, and spatial resources for the coordinated APs, thus opening the door to find ways for improving the overall WLAN performance in terms of both throughput and latency \cite{adame2019time}. Moreover, global scheduling decisions can take into account the needs and requirements of all the APs and stations (STAs) in the vicinity. For instance, real-time streaming can be prioritized by reducing both intra and inter-AP contention.

To support MAPC, the following schemes have been discussed by the IEEE 802.11be Task Group (TGbe) \cite{AdrianGarciaSurvey,EvgenySurvey}: coordinated Orthogonal Frequency Division Multiple Access (c-OFDMA), coordinated Time Division Multiple Access (c-TDMA), coordinated Beamforming (c-BF), coordinated Joint Transmission (c-JT), and coordinated Spatial Reuse (c-SR). Among them, c-TDMA, c-SR, and c-OFDMA are the most suitable candidates to be included in 11be due to their simplicity. Nevertheless, MAPC is still in its infancy, so extensive work must be done to fully understand its potential benefits and drawbacks, design efficient and practical coordination mechanisms and TXOP sharing techniques, and evaluate its performance under realistic conditions with heterogeneous traffic and devices.


\begin{figure*}[t!!!!!]
    \centering
    \includegraphics[scale=0.9]{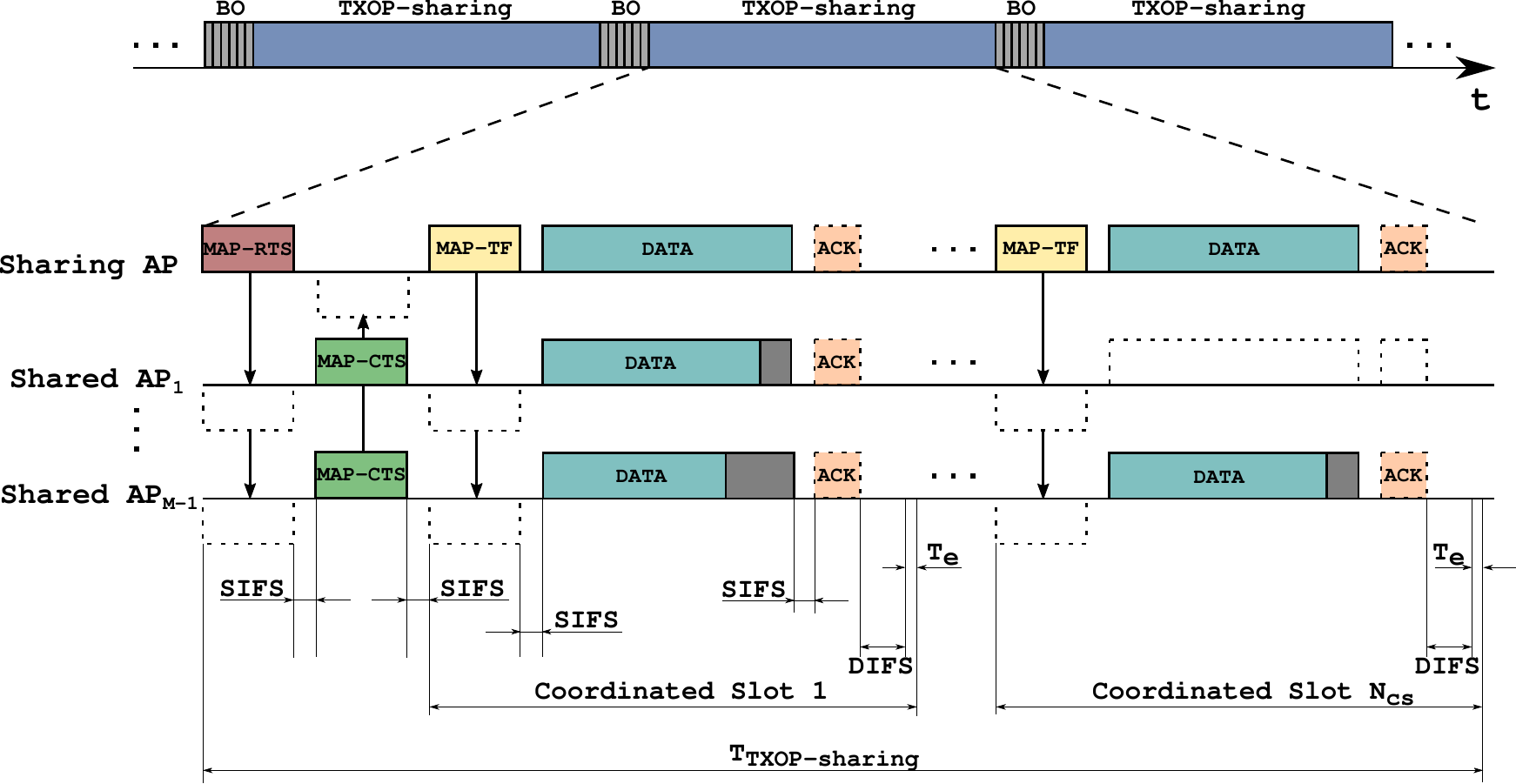}
    \caption{Operation of the proposed c-TDMA/SR scheme. The Sharing AP reserves and sends out the configuration to the Shared APs through MAP-RTS and MAP-TF frames, respectively. The TXOP is divided into coordinated slots, and several APs can transmit simultaneously, depending on the selection made by the Sharing AP.}
    \label{Fig:c-TDMA/SR}
\end{figure*}


In this paper, we show the benefits of TXOP sharing using multi-AP coordination when c-TDMA and c-TDMA/SR schemes are employed. We propose a new MAPC protocol to perform TXOP sharing cooperative transmissions, and a scheduler for the c-TDMA/SR case. The proposed scheduler is able to select the best subset of AP-STA combinations that can transmit concurrently, along with the corresponding power transmission levels. Then, we analyze the performance of c-TDMA and c-TDMA/SR in different scenarios, varying both the number of APs and stations. Results evidence that TXOP sharing effectively reduces the channel contention, and therefore it is able to improve the WLAN performance. However, the required coordination overheads hinder MAPC in terms of throughput when only c-TDMA is employed. This situation is solved by adding SR on top of c-TDMA, which enables throughput gains higher than the 140\%, 80\% and 60\% in the 90\% of the considered scenarios for 4, 3 and 2 overlapping APs, respectively.


\section{Related Work}

Currently, there are only a few works that delve into this topic, and most of the information available comes directly from TGbe documents. In~\cite{MentorConsiderations0590r5}, the authors split the downlink (DL) c-SR procedure into three phases and investigate several operation issues, such as the process to exchange the information about transmission power levels (one-way or bidirectional), path loss, and block acknowledgment. In addition, the works in~\cite{MentorSR1534r1,MentorSR0107r1,MentorSR0576r1} provide simulation results showing the potential performance gains of c-SR. For example, in~\cite{MentorSR1534r1}, the authors show throughput gains for c-SR two times higher than without coordination. Besides, the work in~\cite{MentorSR0107r1} showcases higher throughput when c-SR is compared with the default Enhanced Distributed Channel Access (EDCA) mechanism. Similarly, the authors in \cite{MentorSR0576r1} show the advantages of c-SR compared to c-OFDMA. As a result, the throughput of the former exceeds (twice in some cases) the latter. Finally, a novel transmission scheme for 11be networks, utilizing the concept of multi-AP coordinated OFDMA is proposed in~\cite{WoojinCoOFDMAframework}. The proposed c-OFDMA scheme effectively allows APs to increase the number of transmission opportunities, achieving a higher throughput than DL OFDMA in IEEE 802.11ax.



\section{TXOP sharing with Coordinated Spatial Reuse} \label{section:multi-AP}


To support TXOP sharing between APs, one of the APs acts as the initiator and coordinates a shared transmission, while the other APs will simply follow the received indications. We will refer to the AP initiating the shared transmission as the Sharing AP, and to the rest of APs as the Shared APs. All these APs form a coordination group. Then, the Sharing AP is able to distribute the time it has granted after winning a TXOP with the other APs in the group.


\subsection{Transmission coordination}

The proposed c-TDMA/SR mechanism is shown in Figure~\ref{Fig:c-TDMA/SR}.\footnote{An alternative solution to exchange the information between APs could be a wired backbone, which constitutes the best option in terms of latency and bandwidth.} It allows the AP winning the contention to share its TXOP with other APs in the coordination group. A coordinated transmission starts with the Sharing AP sending a MAP request-to-send (MAP-RTS) frame. If there is no collision, the other APs reply at the same time with a MAP clear-to-send (MAP-CTS) frame.\footnote{As in IEEE 802.11ax, the main purpose of the MAP-CTS is to confirm and relay the channel reservation initiated with the MAP-RTS frame. It is not mandatory to be well-decoded by all stations, although desirable.} At this point, the Sharing AP assumes that all devices in the network have properly set their network allocation vector (NAV), and so the multi-AP transmission will not be disturbed until it ends. 

After the setup, the Sharing AP grants temporal slots, to which we refer as coordinated slots, to the other APs in the coordination group. To do that, the Sharing AP sends a MAP trigger frame (MAP-TF) to allocate the next coordinated slot (i.e., that corresponds to the duration of a transmission) to one or more APs. This frame is also useful for synchronization purposes and it contains a set of configuration parameters, such as maximum physical layer convergence procedure (PLCP) protocol data unit (PPDU) length, coordinated slot duration, total bandwidth, modulation and coding scheme (MCS), and transmission power, that the Shared APs will use in the upcoming transmission. When a coordinated slot is allocated to a single AP, the TXOP is shared following a traditional TDMA scheme. Otherwise, if several APs are allocated to the same slot, TDMA is enhanced with SR.


\subsection{Required coordination information}\label{Section:RSSI_information}


In order to allocate the coordinated slots to different AP-STA links, the Sharing AP needs to know the received signal strength indicator (RSSI) received at the stations from all APs in the coordination group. Therefore, we consider the existence of a sounding stage in which APs request the RSSI values from their associated stations (RSSI values of all APs heard from each station). STAs collect RSSI information from beacon frames, which are transmitted using maximum power, and so it allows to estimate the path loss for all the AP-STA links. This information should be periodically exchanged between all the APs in the group. Accordingly, the Sharing AP employs these RSSI values to decide which AP-STA links are scheduled in the different coordinated slots, the transmission power used by the estimated AP, and the resulting signal-to-interference-plus-noise ratio (SINR) and MCS for each individual AP-STA link.




\subsection{Scheduling Algorithm for the c-TDMA/SR case}\label{Scheduling_subsection}

In c-TDMA/SR, since several APs can transmit in the same coordinated slot, the Sharing AP has to identify which are the best group of suitable AP-STA pairs, determining the transmission power and MCS allocated to each one.

We assume that the Sharing AP knows the average received power at each STA from the different APs. Using that information, the Sharing AP analyzes all the possible combinations of transmitting APs and transmission power values to determine which are the best AP-STA pairs.\footnote{Since the goal of the paper is to understand the potential throughput gains that c-TDMA/SR may bring to Wi-Fi, we have simply applied a brute force approach to systematically explore all possible combinations. The design of a practical low-complexity scheduling algorithm is left as future work.}
For a combination $i$, i.e., $c_{i}$, the Sharing AP estimates its quality using the $\alpha$ parameter as follows: 
\begin{equation}\label{coeff_equation}
\alpha_{i} = \frac{N_{{\rm bits},i}}{T^{\rm c-slot}_{i}},
\end{equation}
where $N_{\text{bits},i}$ is the total number of bits transmitted for the combination $i$, and $T^{\rm c-slot}_{i}$ is the time required to complete all the simultaneous transmissions in combination $i$. Since deterministic packet sizes are considered, the slot duration depends only on the duration of the transmission using the lowest MCS in combination $i$. See Section~\ref{thr_multi-AP_coordination_subsection} for further details.

Once we have computed all $\alpha$ values, we sort accordingly the set of combinations, $C = \{c_{1}, c_{2}, ... c_{K}$\}, where $K$ is the total number of possible combinations, by taking all $\alpha$ values in descending order. Then, we select the subset of best combinations starting from the first one in descending order until all stations associated to APs that belong to the coordination group are scheduled, with the constraint that a station must only be selected once.

We now showcase how the selection is made through an example. Let us have a scenario with two APs and six stations, three associated to AP$_{1}$ (stations 1, 2 and 3) and three associated to AP$_{2}$ (stations 4, 5 and 6). The information about the status of all devices has been collected previously as mentioned before. The Sharing AP is also able to calculate the $\alpha$ coefficients for all combinations. Table~\ref{tab1_combinations} shows an excerpt of the best combinations for this scenario. In this case, $c_{1}$ is the first selected combination. It means that AP$_1$ and AP$_2$ transmit to STA$_1$ and STA$_4$, respectively. Then, the next valid combination is $c_{4}$, because in $c_2$, STA$_4$ appears again, and following the design rules of the scheduler that is not possible. Therefore, $c_{1}$, $c_{4}$ and $c_{9}$ will be used in this toy scenario. Note that in c-TDMA/SR, the Sharing AP can adjust the power level of all APs sharing a coordinated slot to maximize the aggregate throughput. 

\begin{table}
\caption{Example of combinations in a scenario with two APs and six stations.}
\label{tab1_combinations}
\begin{tabular}{|c|c|c|c|c|c|l|}
\hline
                              & \multicolumn{2}{c|}{AP$_{1}$}                         & \multicolumn{2}{c|}{AP$_{2}$}                         & \multicolumn{2}{c|}{}                                   \\ \cline{2-5}
\multirow{-2}{*}{Combinations} & P {[}dBm{]}          & STA                      & P {[}dBm{]}          & STA                      & \multicolumn{2}{c|}{\multirow{-2}{*}{$\alpha$
 {[}Mbps{]}}} \\ \hline
c$_{1}$                             & 23                   & 1 & 23                    &  4 & \multicolumn{2}{c|}{80.62}                              \\ \hline
c$_{2}$       &      23              & 3 & 23                   & 4  & \multicolumn{2}{c|}{80.62}                            \\ \hline
\multicolumn{1}{|c|}{\vdots}         & \multicolumn{1}{c|}{\vdots} & \multicolumn{1}{c|}{\vdots}    & \multicolumn{1}{c|}{\vdots} & \multicolumn{1}{c|}{\vdots}    & \multicolumn{2}{c|}{\vdots}                                   \\ \hline
c$_{4}$                              & 20                    &  3 & 20                   &  5 & \multicolumn{2}{c|}{80.62}                              \\  \hline
\multicolumn{1}{|c|}{\vdots}         & \multicolumn{1}{c|}{\vdots} & \multicolumn{1}{c|}{\vdots}    & \multicolumn{1}{c|}{\vdots} & \multicolumn{1}{c|}{\vdots}    & \multicolumn{2}{c|}{\vdots}                                   \\ \hline
c$_{9}$                              & 23                    & 2 & 20                    &  6 & \multicolumn{2}{c|}{74.81}                              \\ \hline
 \vdots                              &            \vdots           &              \vdots            &           \vdots            &                          & \multicolumn{2}{c|}{\vdots}                                   \\ \hline
\end{tabular}
\end{table}


\section{System model}\label{system_model}

\begin{figure}[t!!!]
    \centerline{\includegraphics[scale=0.575]{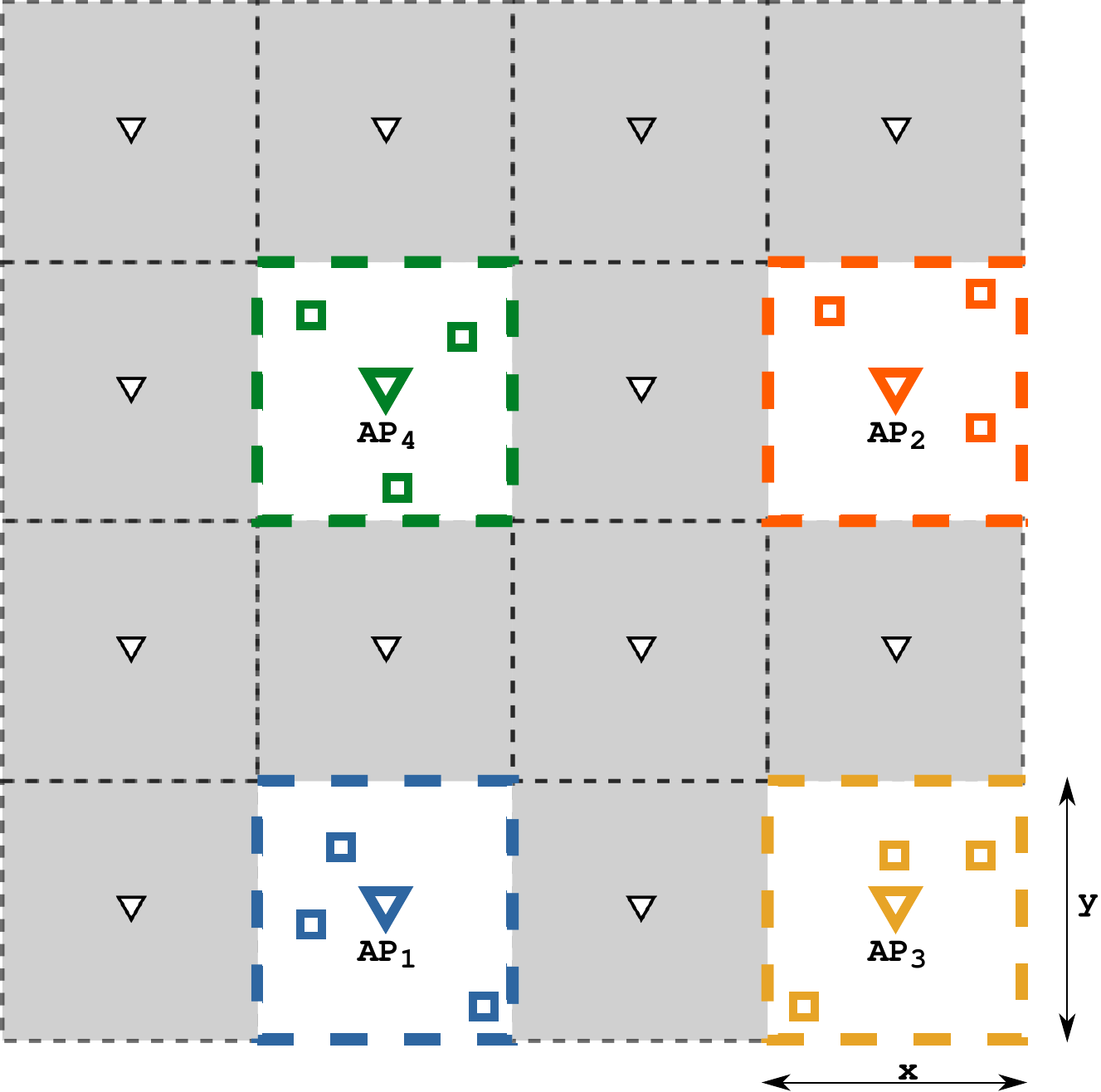}}
    \caption{An Enterprise WLAN scenario, with multiple APs, but only four using the same channel, and so overlaping.}
    \label{Fig:MAPC_scenarios} 
\end{figure}

We consider a dense enterprise WLAN such as the one depicted in Fig.~\ref{Fig:MAPC_scenarios}. However, we focus only on the group of $M$ APs that share the same channel (colored APs). Each one of the $M$ APs is deployed to cover an area of $A=x\cdot y$ m$^2$. $N_m$ stations are then deployed uniformly at random in each area $A_m$, and so they associate to the corresponding AP$_m$. We assume that all $M$ APs and $N=N_m\cdot M$ STAs are within the coverage area of the others, and therefore all of them can correctly receive and decode any transmitted frame in the WLAN.





Multiple transmission rates are allowed, so stations close (far) from the AP use higher (lower) MCSs to transmit and receive data. The MCS used by the corresponding AP $m$ to transmit to station $n$ depends on the received SINR (which is estimated using the collected RSSI information, see Section~\ref{Section:RSSI_information}). To allocate a specific MCS to a station, we employ the mechanism presented in \cite{EvgenyMCS}, which defines the SINR ranges corresponding to each MCS so that a low transmission error is guaranteed.

APs are allowed to use different transmission power levels. The clear channel assessment (CCA) / preamble detection (PD) is set to the energy detection (ED) threshold. Full-buffer downlink traffic is considered, and APs transmit with the same probability to all the associated stations. 

The path loss effects are modelled using the TGax model for Enterprise Scenarios \cite{pathloss}:
\begin{equation}\label{pl_equation}
P_{L} = 40.05 + 20\log_{10}\left(\frac{\min(d,B_{p})f_{c}}{2.4}\right) + P' + 7W_{n} \text{,}
\end{equation}
where $d$ is the distance between the transmitter and the receiver in meters, $f_{c}$ is the central frequency in GHz, $W_{n}$ is the number of walls and $P'$ is given by $P' = 35\log_{10}(d/B_{p})$,
when $d$ is higher than the breaking point $B_{p}$. Otherwise, it is zero.


The following two cases have been defined to compare the use (or not) of TXOP sharing coordination mechanisms. DCF with the RTS/CTS mechanism is employed as channel access in all cases.

\begin{enumerate}
%


\item \textbf{Multi-AP without coordination (nc-MAP)}: The $M$ APs contend to access the channel, using the regular DCF mechanism, which potentially results in collisions when nodes select the same backoff counter. 



\item \textbf{Multi-AP coordination}: The $M$ APs belong to the same multi-AP coordination group. Thus, they can coordinate their transmissions following the indications of the Sharing AP.

\end{enumerate}


\section{Throughput Analysis}

To compute the throughput achieved by the stations we resort to Bianchi's IEEE 802.11 throughput model \cite{bianchi2000performance}. Thus, the aggregate WLAN throughput corresponds to the sum of the throughput achieved by all APs, and it is given by
\begin{equation}\label{S_TDMASR_equation}
S = \frac{p_{s} N_{\rm bits}}{\E[T]} \text{,} 
\end{equation}
where $p_s$ is the probability that a backoff slot contains a successful transmission, $N_\text{bits}$ is the total number of bits included in a successful transmission, and $\E[T]$ is the average duration of backoff slot following Bianchi's model nomenclature. In the following, we will see how to compute all these parameters for each case.






\subsection{Multi-AP without coordination} 

In this first case, there are $M$ APs contending for accessing to the channel using DCF, and therefore collisions between them appear. Let $N_m$ be the number of stations associated to AP $m$. The parameters of \eqref{S_TDMASR_equation} are $N_{\rm bits} = L$, where $L$ is the number of bits sent in each single frame, $p_{s} = M\tau(1-\tau)^{M-1}$ and $ \E[T]  = p_{e} T_{e} + \sum_{m=1}^{M}{\sum_{n=1}^{N_m}{w_{m,n} T_{m,n}}} + p_{c} T_{c} $, where $p_{e}=(1-\tau)^{M}$ and $T_{e}$ are the probability and the duration of an empty slot, respectively; $p_{c} = 1-p_{e}-p_{s}$ is the probability to observe a collision slot with duration $T_c$, given by
\begin{align}
    T_{c} &=  T_{\rm RTS} + T_{\rm CTS-TO} \text{,} \nonumber
\end{align}
where $T_{\rm CTS-TO}$ corresponds to the CTS timeout.
Now, $w_{m,n}$ is the probability that the AP $m$ transmits successfully, divided by the number of stations it has associated, i.e., $w_{m,n}  = \frac{\tau(1-\tau)^{M-1}}{N_m}$. Besides, we define $T_{m,n}$ as the duration of a successful slot including a transmission to station $n$ from AP $m$. It depends on the transmission rate $R_{m,n}$ used and so to which station the packet is directed, thus it yields
\begin{align}
    T_{m,n} & =T_{\rm RTS} + T_{\rm SIFS} + T_{\rm CTS} +  T_{\rm SIFS} +  \nonumber \\
    & + T_{\rm DATA}(R_{m,n}) + T_{\rm SIFS} + T_{\rm ACK} + T_{\rm DIFS} + T_{e} \text{.}
\end{align}

\subsection{Multi-AP coordination}\label{thr_multi-AP_coordination_subsection}

The following analysis is valid for both c-TDMA and c-TDMA/SR, unless otherwise specified. First, we assume that only the Sharing AP is allowed to initiate a transmission, and therefore, there are no collisions. In each TXOP, a single packet is transmitted to each station associated to an AP of the coordination group. For example, with $N=6$ stations, a c-TDMA transmission includes six coordinated slots. In the c-TDMA/SR case, the number of coordinated slots can be lower, as they can include simultaneous transmissions to different stations.

In this case, therefore, to compute the WLAN throughput using \eqref{S_TDMASR_equation}, the different parameters are $N_{\rm bits} = N L$, $p_e = 1-\tau$, $p_s=\tau$, $p_c = 0$, and $\E[T] = p_e T_e + p_{s}T_{\rm TXOP-sharing} $, where $T_{\rm TXOP-sharing}$ is either the duration of a c-TDMA or a c-TDMA/SR transmission depending on which operation mode is considered. It is given by
\begin{align}\label{tcycle_equation}
T_{\rm TXOP-sharing} &= T_{\rm MAP-RTS} + T_{\rm SIFS} + T_{\rm MAP-CTS} + \nonumber \\  &+ T_{\rm SIFS} +  \sum_{i=1}^{N_{\rm cs}}{T^{\rm{c-slot}}_{i}}   \text{,} \nonumber
\end{align}
where $N_{\rm cs}$ is the number of c-TDMA coordinated slots, and $T^{\rm{c-slot}}_{i}$ the duration of the $i$th coordinated slot. The duration of $T^{\rm{c-slot}}_i$ depends on the individual duration of the transmissions scheduled in it, and corresponds to the duration of the longest one, i.e.,
\begin{align}
    T^{\rm{c-slot}}_{i}=T_{\rm MAP-TF} + T_{\rm SIFS} + \max\{\mathcal{T}_{i}\}, \nonumber
\end{align}
where $T_{\rm MAP-TF}$ is the duration of the trigger frame sent by the Sharing AP to indicate the allocated AP-STA links, and $\mathcal{T}_{i}$ is the set containing the duration of each individual transmission in the coordinated slot $i$. The duration of an individual transmission from AP $m$ to STA $n$ is given by
\begin{equation}\label{ts_equation}
T_{m,n} =  T_{\rm DATA}(R_{m,n}) + T_{\rm SIFS} + T_{\rm ACK} + T_{\rm DIFS}.
\end{equation}
Note that, differently from the previous case, we do not include the RTS/CTS in each transmission since we consider that the channel is already reserved for the c-TDMA/SR transmission after the initial MAP-RTS/MAP-CTS exchange. 

\begin{figure*}[t!]
    \centering
    \begin{subfigure}[b]{0.3\textwidth}
        \centering
        \includegraphics[width=\textwidth]{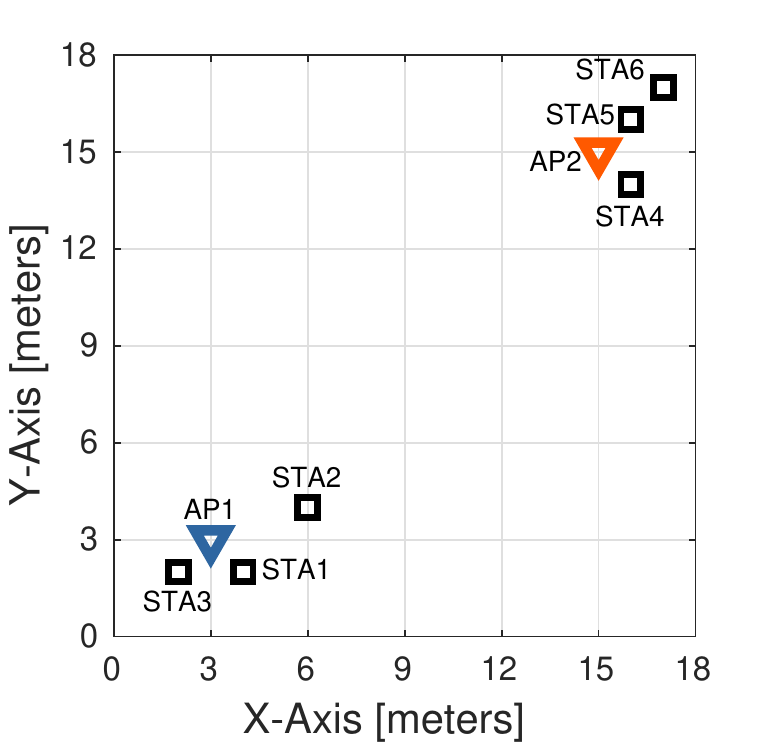}
        \caption{Toy scenario 1 (2 APs vs 6 STAs).}
        \label{toy_scenario1}
    \end{subfigure}
    \hfill
    \begin{subfigure}[b]{0.3\textwidth} 
        \centering
        \includegraphics[width=\textwidth]{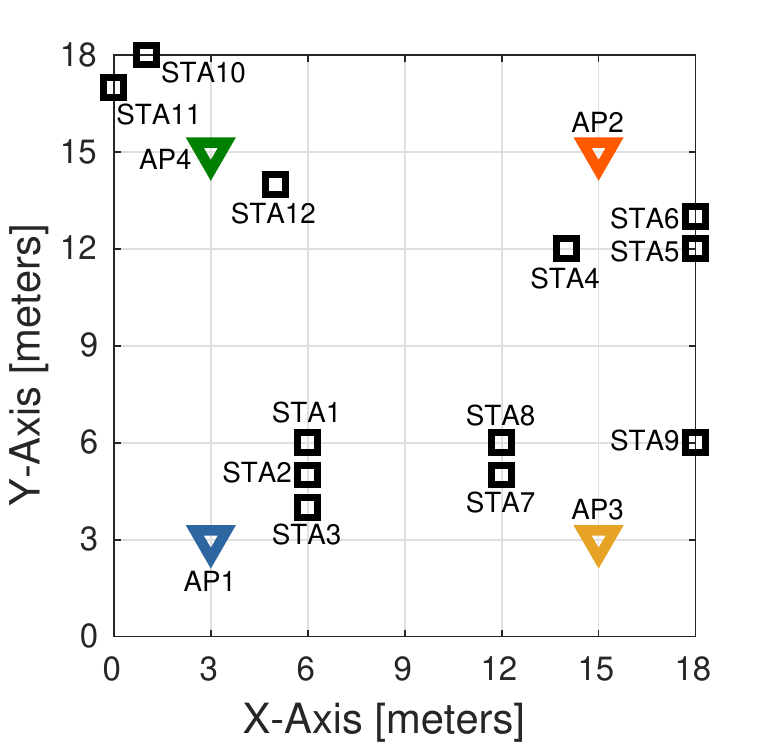}
        \caption{Toy scenario 2 (4 APs vs 12 STAs).}
        \label{toy_scenario2}
    \end{subfigure}
    \hfill
    \begin{subfigure}[b]{0.3\textwidth}
        \centering
        \includegraphics[width=\textwidth]{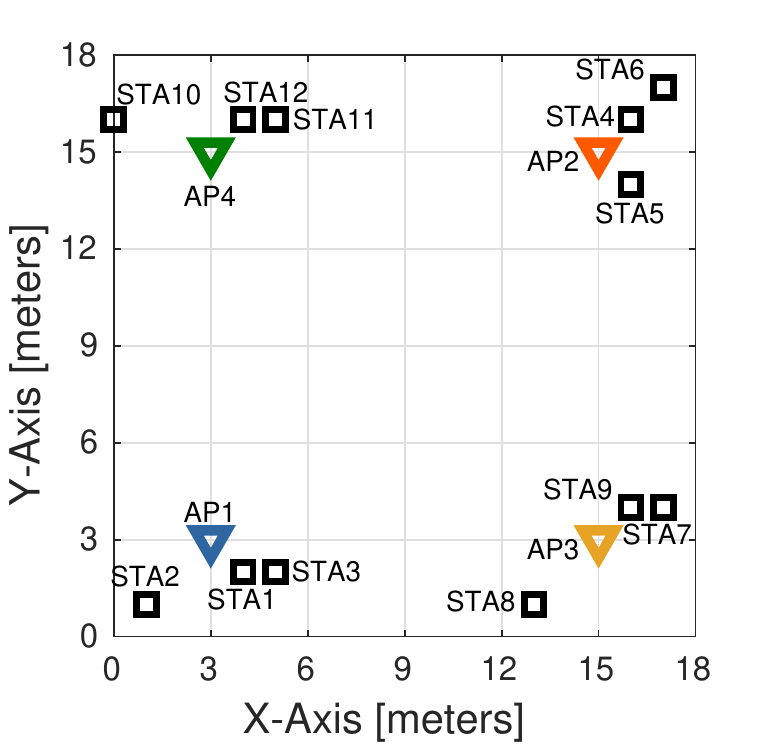}
        \caption{Toy scenario 3 (4 APs vs 12 STAs).}
        \label{toy_scenario3}
    \end{subfigure}
    \newline
    
    \begin{subfigure}[b]{0.3\textwidth}
        \centering
        \includegraphics[width=\textwidth]{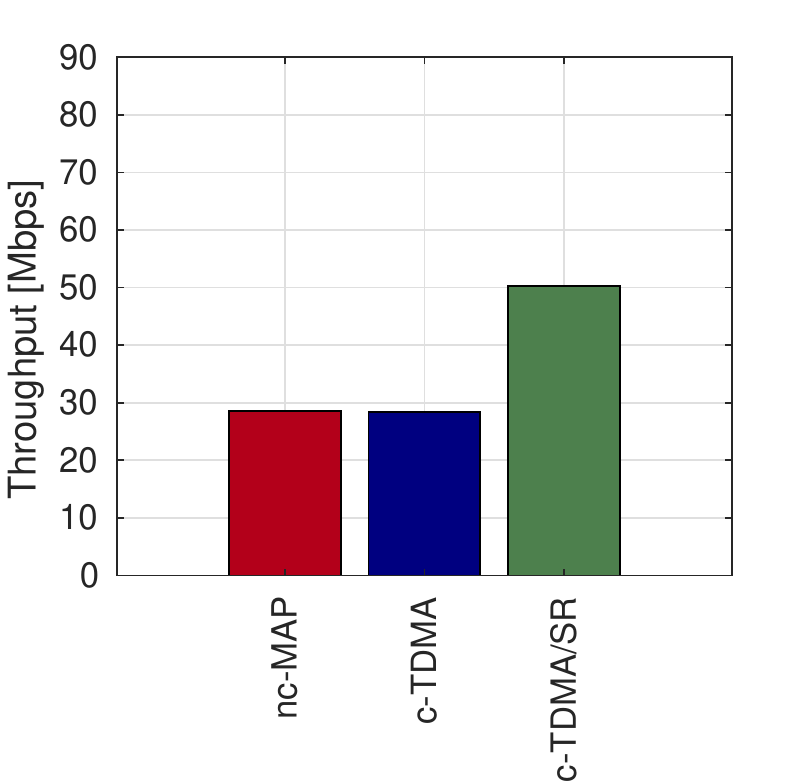}
        \caption{Throughput in Toy scenario 1.}
        \label{throughput_toy_scenario1}
    \end{subfigure}
    \hfill
    \begin{subfigure}[b]{0.3\textwidth} 
        \centering
        \includegraphics[width=\textwidth]{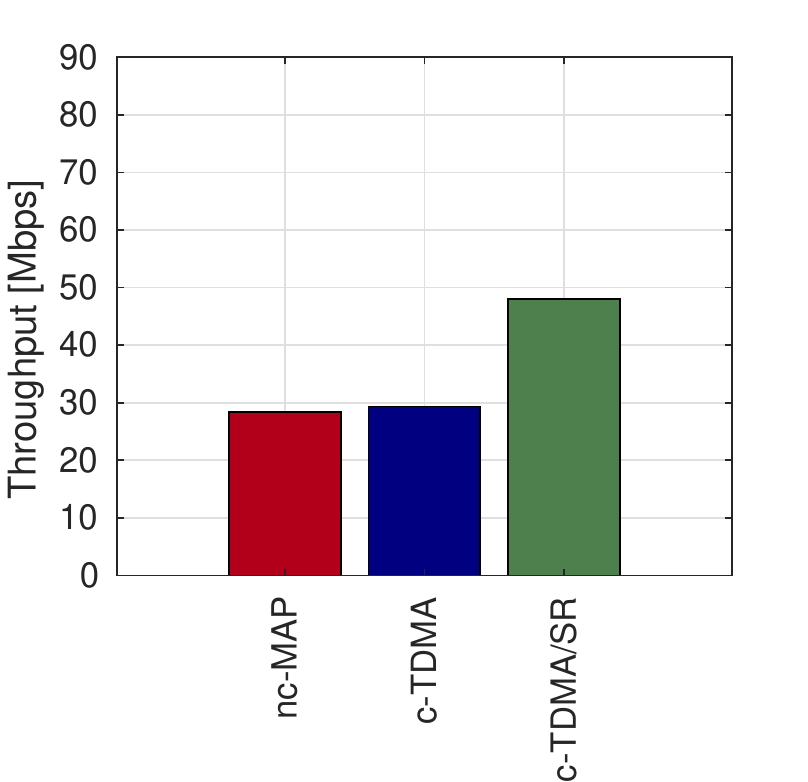}
        \caption{Throughput in Toy scenario 2.}
        \label{throughput_toy_scenario2}
    \end{subfigure}
    \hfill
    \begin{subfigure}[b]{0.3\textwidth}
        \centering
        \includegraphics[width=\textwidth]{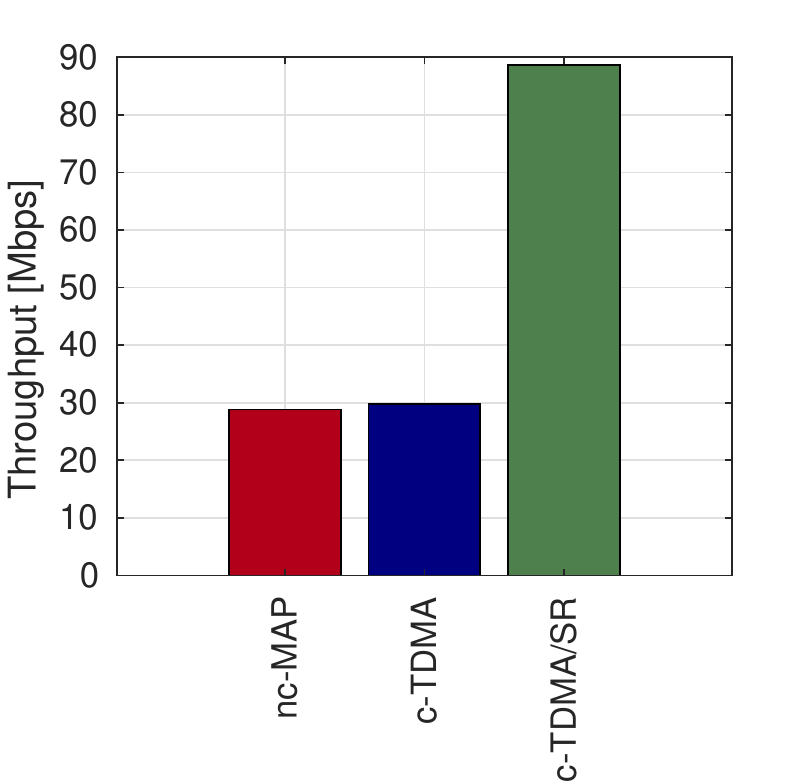}
        \caption{Throughput in Toy scenario 3.}
        \label{throughput_toy_scenario3}
    \end{subfigure}    
    \caption{Three different toy scenarios and the corresponding throughput for them.}
    \label{fig:Toy_Scenarios}
\end{figure*}


\section{Results}

In this section, we present and discuss the results obtained from both specific toy scenarios and randomly generated scenarios. In particular, considering the scenario depicted in Fig.~\ref{Fig:MAPC_scenarios} from Section~\ref{system_model}, we have designated AP$_{1}$ as the Sharing AP, and all APs are deployed at the center of each 6$\times$6 meters room and use the same 80~MHz channel. The same number of stations, deploying them uniformly at random, is placed in each room. Besides, results were obtained using the default path-loss parameters proposed for IEEE 802.11ax, i.e., $B_{p} = 10$ and $W_{n} = 3$.

Also, for simplicity, we consider that APs implement a single backoff stage, and so the transmission probability in Bianchi's model is computed as $\tau = {2}/{(\text{CW}_\text{min} + 2)}$, where CW$_{\rm min}$ is the minimum contention window. The parameters for the numerical simulations considered in this paper are shown in Table~\ref{tab2_simulation_paramters}.

\begin{table}
    \caption{Simulation Parameters.}
    \label{tab2_simulation_paramters}
    \begin{center}
        \begin{tabular}{|c|c|}
        \hline
        \textbf{Parameter} & \textbf{Value} \\
        \hline
        AP Tx-power values [dBm] & 11, 14, 17, 20, 23 \\
        \hline
        CCA [dBm] & -82 \\
        \hline
        11ax MCS [index] & 0-10 \\
        \hline
        $L$ [bytes] & 1500 \\
        \hline
        Number of spatial streams & 1  \\
        \hline
        Legacy preamble [$\mu$s] & 20 \\
        \hline
        OFDM symbol duration [$\mu$s] & 12.8 \\
        \hline
        Guard interval duration [$\mu$s] & 0.8 \\
        \hline
        $T_{\text{MAP-RTS}}$ [$\mu$s] & 80\\
        \hline
        $T_{\text{MAP-CTS}}$ [$\mu$s] & 62 \\
        \hline
        $T_{\rm CTS-TO}$ [$\mu$s] & 41 \\
        \hline
        $T_{\rm MAP-TF}$ [$\mu$s] &  76\\
        \hline
        T$_{e}$ [$\mu$s] &  9 \\
        \hline
        CW$_\text{min}$ & 15 \\
        \hline
        \end{tabular}
    \end{center}
\end{table}


\subsection{Toy scenarios}

In this section, we present the results for nc-MAP, c-TDMA, and c-TDMA/SR operation modes in three specific deployments. Our goal is to illustrate the dependence of the achievable throughput with the specific deployment.

Figure~\ref{fig:Toy_Scenarios} depicts three toy scenarios with a different number of APs and three stations associated to each of them, as well as the corresponding throughput for the aforementioned scenarios. In all toy scenarios, the c-TDMA throughput is similar to the nc-MAP. Note that, while in the scenarios with 2 APs the c-TDMA throughput is sometimes slightly lower due to the coordination overheads, when more APs are added to the coordination group, these overheads get compensated by the reduced channel contention. Besides, in all toy scenarios, the throughput for c-TDMA/SR exceeds the c-TDMA throughput by a factor between 1.6 and 3. 

In Toy scenario 1, we can observe that c-TDMA/SR achieves better throughput than c-TDMA because the stations are close to their corresponding APs, and so it is possible to perform simultaneous SR transmissions by adjusting the transmission power levels of the two APs. Then, enabling SR allows transmitting the same number of packets in a TXOP (one to each station) in a shorter time, so increasing the throughput. This scenario represents the example shown in Section~\ref{Scheduling_subsection}, so the transmission power levels, as well as the best subset of combinations selected by the Sharing AP, are exhibited in Table~\ref{tab1_combinations}.

Unlike Toy scenario 1, in Fig.~\ref{toy_scenario2}, all stations are widely dispersed throughout each sub-area (some of them are on the limits of each room and close to other APs), which leads to a higher level of interference at the receivers in case several APs transmit at the same time. Therefore, simultaneous transmissions (c-TDMA/SR) are not as advantageous as in other scenarios, so the maximum gain cannot be achieved.

Finally, Fig.~\ref{toy_scenario3} shows the toy scenario that results in the highest throughput for c-TDMA/SR. All devices are placed close to their respective corners and, therefore, they will experience low interference from other ongoing transmissions, and APs will be able to transmit using high MCS indexes. All APs can also transmit in each coordinated slot, reducing the airtime spent to send a packet to all stations.


\subsection{Randomly generated Scenarios}

Results in this section were obtained through the simulation of 10,000 different deployments. In each deployment, while the APs are kept at the center of each room, the positions of the stations are generated uniformly at random.

\begin{figure}
    \centerline{\includegraphics[scale=0.6]{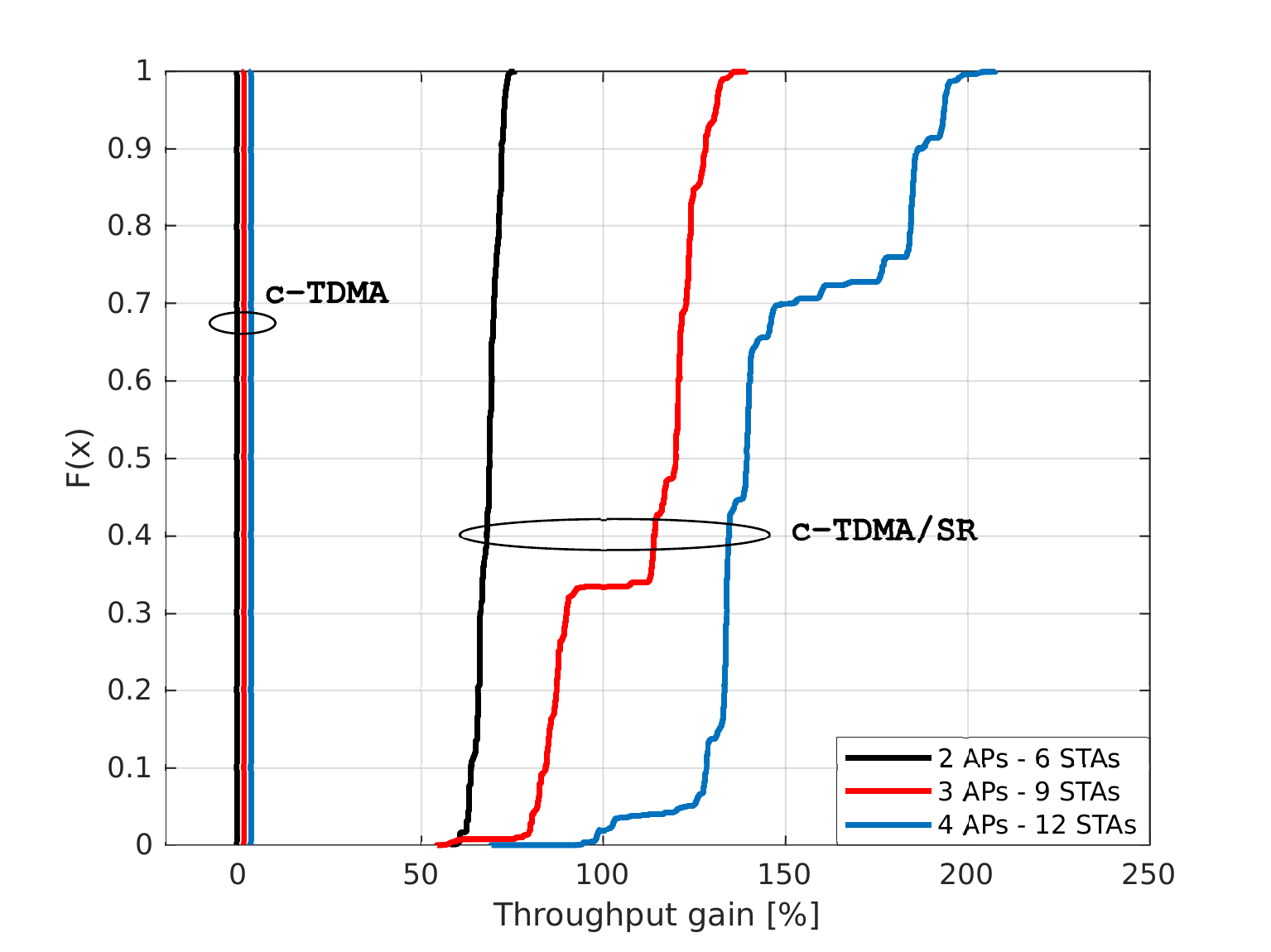}}
    \caption{Throughput gain of c-TDMA and c-TDMA/SR vs nc-MAP. CDF for randomly generated scenarios with 2, 3 and 4 transmitting APs, and 6, 9 and 12 stations, respectively.}
    \label{random_scenarios_cdf_cTDMASR_vs_no_cMAP}
\end{figure}

Figure \ref{random_scenarios_cdf_cTDMASR_vs_no_cMAP} shows the Cumulative Distribution Function (CDF) of the throughput gain (in percentage) between c-TDMA and c-TDMA/SR operation modes with respect to nc-MAP. It can be observed that c-TDMA does not significantly improve the throughput as the maximum gain is close to 5\%. We can also highlight that the difference between c-TDMA/SR and c-TDMA modes is significant, and it can be greater than 190\% in the 10\% of the four-AP scenarios, and 140\% in the 90\% of the cases. Additionally, the minimum throughput gain is around 60\% for a few scenarios, as observed in Toy scenario 2 from the previous section.  

\begin{figure}
    \centerline{\includegraphics[scale=0.6]{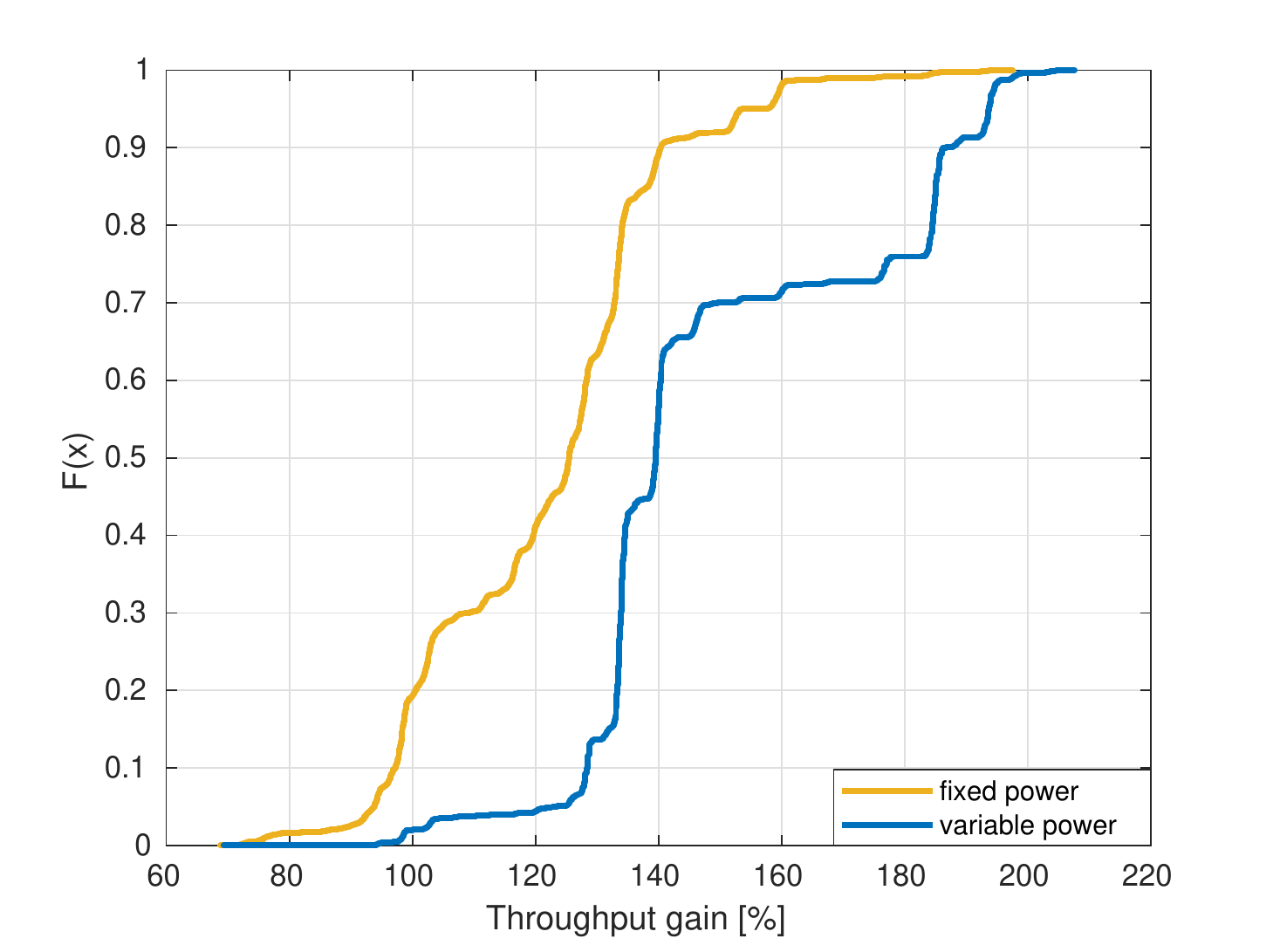}}
    \caption{Throughput gain of c-TDMA/SR with fixed and variable power vs nc-MAP. CDF for randomly generated scenarios with 4 transmitting APs and 12 stations.}
    \label{random_scenarios_cdf_cTDMASR_fixed_variable_power}
\end{figure}

Moreover, Fig.~\ref{random_scenarios_cdf_cTDMASR_fixed_variable_power} shows the CDF of the throughput gain (in percentage) of c-TDMA/SR with respect to nc-MAP operation mode in two cases: \emph{a)} fixed transmission power, and \emph{b)} variable transmission power. The transmission power is set to 23 dBm for the fixed case. A higher gain when variable power is employed for c-TDMA/SR operation mode is observed, allowing APs to adjust the transmission power to more convenient levels. This figure also shows that in many cases, reducing the transmission power and so using lower MCS values, is compensated by the fact that there is less interference, and so, the overall transmission rates of the network are better. Thus, in 5\% and 95\% of the cases, the throughput gain achieved is 30\% greater with dynamic transmission power.

\begin{figure}
    \centerline{\includegraphics[scale=0.6]{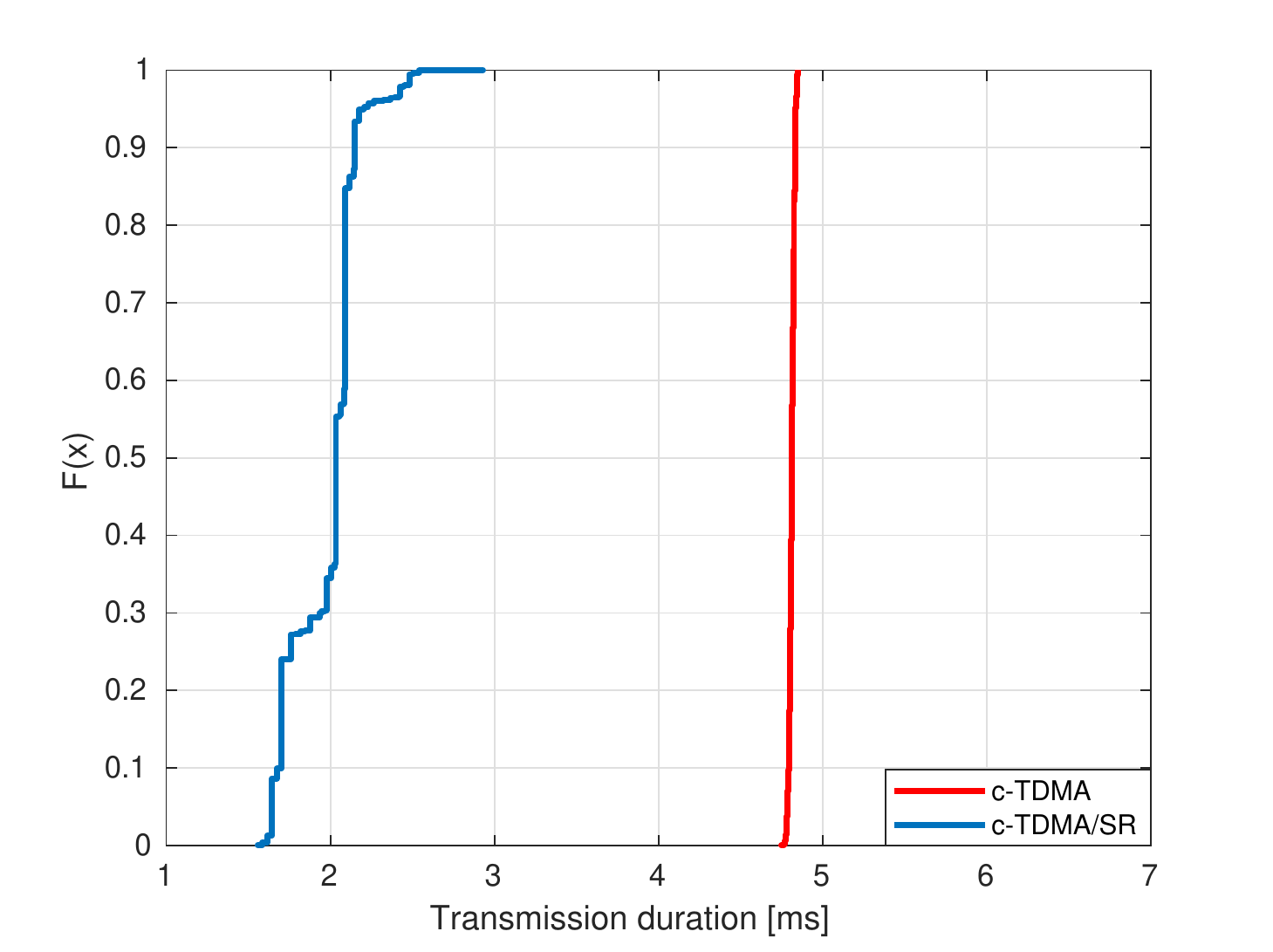}}
    \caption{CDFs of the transmission duration for c-TDMA and c-TDMA/SR systems. Simulation obtained using 4 APs, and 12 stations}
    \label{c-TDMA_c-TDMASR_tx_duration}
\end{figure}

In addition, Fig.~\ref{c-TDMA_c-TDMASR_tx_duration} shows the CDFs of the transmission duration for c-TDMA/SR and c-TDMA operation modes. It can be observed that although coordinated slots could be longer due to lower MCSs when using the former, a higher number of APs transmitting in parallel reduces the number of coordinated slots, the transmission duration and thus the average TXOP. Finally, even though we have not considered packet aggregation in this work, it is expected that by transmitting multiple packets to each STA, we can further improve c-TDMA/SR compared to c-TDMA throughput.

\section{Conclusions}

In this paper, we have explored the benefits of TXOP sharing multi-AP coordination for next-generation WLANs. We proposed a MAPC packet exchange and a scheduling algorithm to select the best AP-STAs pairs for each coordinated slot in the c-TDMA/SR scheme. The simulation results show that c-TDMA can potentially reduce contention by sharing the TXOP, but it does not necessarily improve the system throughput, which is solved using spatial reuse. Although the success of c-TDMA/SR depends on the device locations, we foresee higher gains in a scenario where multiple transmitting APs can dynamically adjust their transmission power. 

Since our packet exchange model has been designed to obtain results that validate the feasibility of using coordinate schemes, it has to be properly improved to not only consider transmission overheads, but also the information between APs in the previous stages, i.e., group formation, sounding process, etc. Moreover, future work should also include the design and implementation of more practical, low-complexity scheduling algorithms, the evaluation of larger and dynamic scenarios, the combination of TXOP sharing with other technologies such as MU-MIMO and OFDMA, and the inclusion of realistic traffic profiles for both downlink and uplink. Special interest will be also placed in the performance of TXOP sharing strategies in terms of latency.



\section*{Acknowledgements}

The work of D. Nuñez and B. Bellalta was supported by WINDMAL PGC2018-099959-B-I00 (MCIU/AEI/FEDER,UE) and Cisco.

The authors also want to acknowledge the comments and suggestions received during the review process.

\bibliographystyle{./bibliography/IEEEtran}
\bibliography{main}

\end{document}